\documentstyle[aps,preprint,epsf]{revtex}
\tighten

\def\ev{\,{\rm eV}}
\def\gev{\,{\rm GeV}}

\def\rmd{{\rm d}}

\def\etal{{\it et al.}}
\def\ibid{{\it ibid.}}

\def\A{{\rm A}}
\def\baryon{{\rm baryon}}

\def\F{{\rm F}}
\def\halo{{\rm halo}}
\def\GZK{{\rm GZK}}
\def\R{{\rm R}}
\def\SM{{\rm SM}}
\def\SUSY{{\rm SUSY}}

\begin{document}
\draft 
\small
\preprint{OUTP-01-20-P}
\title{The high energy cosmic ray spectrum from relic particle decay}
\author{\large Subir Sarkar\thanks{sarkar@thphys.ox.ac.uk} and 
         Ramon Toldr\`a\thanks{toldra@thphys.ox.ac.uk}}
\address{Theoretical Physics, University of Oxford,
          1 Keble Road, Oxford OX1 3NP, UK}
\date{submitted to {\sf Nucl. Phys. B}}
\maketitle
\begin{abstract}
It has been speculated that the recently detected ultra-high energy
cosmic rays may originate from the decays of relic particles with mass
of order $10^{12}$ GeV clustered in the halo of our Galaxy. This
hypothesis can be tested through forthcoming measurements of the
spectra of both high energy cosmic nucleons and neutrinos, which are
determined in this model by the physics of QCD fragmentation, with no
astrophysical uncertainties. We evolve fragmentation spectra measured
at LEP energies up to the scale of the decaying particle mass by
numerical solution of the DGLAP equations. This enables incorporation
of the effects of supersymmetry on the development of the cascade and
we also allow for decays into many-particle states. The calculated
spectral shape agrees well with present cosmic ray data beyond the
Greisen-Zatsepin-Kuzmin energy.
\end{abstract}

\section{Introduction}
\label{intro}
Within the last decade, the Fly's Eye \cite{fe} and its successor
HiRes \cite{hires}, as well as the Akeno Giant Air Shower Array
(AGASA) \cite{agasa}, have detected over 70 ultra-high energy cosmic
rays (UHECRs) with energies measured reliably to be in excess of
$E_\GZK\simeq4\times10^{19}\ev$ --- the Greisen-Zatsepin-Kuzmin (GZK)
`cutoff' energy \cite{gzk}. The older air shower arrays at Volcano
Ranch \cite{vr}, Haverah Park \cite{hp}, and Yakutsk \cite{yak} have
added another 45 (recalibrated) post-GZK events. At energies above
$E_\GZK$, the typical range of protons falls rapidly because of
interactions with cosmic microwave background photons
\cite{propag,ac96}, becoming as low as $\sim20$~Mpc at
$3.2\pm0.9\times10^{20}\ev$, the highest energy event observed by
Fly's Eye; the range is even smaller for heavy nuclei
\cite{nuclei}.\footnote{The shower elongation rate measured by Fly's
Eye \cite{fe,fecomp} and HiRes \cite{hirescomp} indicates a change in
composition from heavier nuclei towards nucleons at the highest
energies. Muon data from AGASA \cite{agasacomp} are consistent with
this trend when analysed using the same hadronic interaction model
\cite{dms98,nhsik00}.} For photons with energy above $E_\GZK$, the
dominant opacity is due to scattering on the extragalactic radio
background which is poorly known experimentally \cite{radio}; the mean
free path is estimated to be $\sim1-5$~Mpc at $10^{21}\ev$
\cite{photon}.\footnote{The highest energy event \cite{fe} had a depth
of maximum of $815^{+45}_{-35}$~g\,cm$^{-2}$ cf.  the expected value
of $1075$~g\,cm$^{-2}$ for a photon primary \cite{hvsv95}. The muon
content of horizontal showers in the Haverah Park data implies an
upper bound of 65\% on the photon component, and a (hadronic
interaction model dependent) upper bound of $\sim50\%$ on the iron
nuclei content, of post-GZK UHECRs \cite{ahvwz00}. It should be
emphasised however that a complete understanding of air shower
development is still lacking, so the composition of UHECRs is not
definitively established yet \cite{airshower,heavy}.} There is no
restrictive constraint on the range if the UHECR primaries are neutral
and weakly interacting, viz. neutrinos which have only a tiny
probability for scattering on the relic cosmic neutrino background
\cite{neutrino,decay}.\footnote{However if the background neutrinos
have mass of $\sim0.1-1\,\ev$, then UHE cosmic neutrinos can
annihilate resonantly on these to generate `Z-bursts' which create
photons and nucleons above $E_\GZK$ \cite{zburst}. To match the
observed UHECR flux given experimental limits on the UHE cosmic
neutrino flux \cite{zburstnot} requires however a substantial increase
in the local density of relic neutrinos \cite{zburstprobs}.} This also
means however that neutrinos cannot initiate the observed airshowers
as their cross-section for deep inelastic scattering on nucleons is
far too small \cite{nu}.\footnote{Neutrinos may interact more strongly
if dramatic new physics is invoked e.g. extra dimensions and quantum
gravity at the TeV scale \cite{nustrong}. However explicit
calculations indicate so far that the increased cross-section would
still be inadequate to create the observed air-showers
\cite{nunot}. This possibility is testable through studies of
penetrating airshowers \cite{nustrongtest} and upward-going muons
\cite{halzen}.}  Thus all the available evidence suggests that cosmic
rays with energies $E>E_\GZK$ are protons, in which case they must
originate within the Local Supercluster.\footnote{A suggested
correlation between UHECR arrival directions and cosmologically
distant radio sources \cite{fr} motivates the possibility that the
primary is a new heavy hadron (e.g. a supersymmetric $uds$-gluino
state) with a higher GZK cutoff \cite{uhecron}. The correlation has
however been questioned \cite{nocorr}; moreover there are stringent
experimental bounds on new stable strongly interacting particles
\cite{rpp}.}

However they cannot originate in the Galactic disk since significant
anisotropies would then be expected at post-GZK energies
\cite{galaniso}, while the observed sky distribution is consistent
with isotropy \cite{anisofe,anisoagasa}.\footnote{At around
$10^{18}\ev$ Fly's Eye \cite{anisofe} and AGASA \cite{diskaniso} have
detected anisotropy towards the Galactic plane, consistent with the
usual belief that cosmic rays at these energies have a galactic
origin.}  A number of doublet and triplet events, collimated within
the experimental angular resolution of a few degrees, have been
observed but the chance probability of these arising from an isotropic
distribution is estimated to be of order 10\% \cite{cluster}, hence
statistically not significant.\footnote{However if only the paired
events within $\pm10^0$ of the supergalactic plane are considered, the
chance probability that they arise from an isotropic distribution is
less than 1\% \cite{cluster}.}

There have been a plethora of suggestions as to the possible origin of
UHECRs, which may be broadly classified into `top-down' or `bottom-up'
models. In the former class, the extreme energies are supposedly
provided by the decay of relic topological defects (TD)
\cite{Hill,td,bbv98,tdgamnu} or super-massive particles
\cite{shdm,chor,bkv97,bs98}. In the latter class, the UHECRs are
assumed to have been accelerated up from low energies in astrophysical
sites such as gamma-ray bursts \cite{grb}, active galaxies
\cite{radiogal} or super-massive black holes in dormant quasars
\cite{bg99}. In fact there are severe difficulties with accelerating
particles to the highest observed energies by {\em any} known
astrophysical mechanism \cite{astro}. Moreover even if such sources
are distributed nearby like the observed galaxies, the injection
spectrum must be harder than $\rmd\!N/\rmd\!E\propto\,E^{-2}$ in order
to overcome the GZK energy losses and match the measured UHECR flux
\cite{ac96,injection}, so the energetic demands are
considerable. Gamma-ray bursters may possibly have the required energy
but, being more abundant at high redshift, cannot yield the locally
observed UHECR flux \cite{grbnot}. There are a few active galaxies
within the `GZK sphere' but even if they can accelerate UHECR these
would need to be deflected through rather large angles (and
isotropised) by the intergalactic magnetic field \cite{es95}; this
requires the field strength to be $\sim10^3$ times stronger than the
upper limit of order nanogauss usually inferred from observational
bounds on Faraday rotation in distant radio sources \cite{igmf}. There
is ongoing discussion concerning the possibility of such high fields
\cite{himag}, as well as of specific sources such as Cen A
\cite{cena,fp00,ils01} and Virgo A \cite{m87,m87not}. Although there
is no compelling astrophysical model for post-GZK UHECRs, some of
these suggestions are interesting in their own right and can be tested
with forthcoming data.

In this paper, we investigate the spectral signature of `top-down'
models. Relic topological defects will in general be cosmologically
distant and release particles at the GUT-scale, resulting in a
typically excessive flux of secondary low energy $\gamma$-rays as the
primaries interact with intergalactic radiation fields
\cite{tdnot,tdgamnu}. So we favour the possibility that UHECRs
originate from the decay of super-massive relic particles which, being
cold dark matter (CDM), are {\em locally} clustered in the halo of our
Galaxy, thus naturally evading the GZK-cutoff and also presenting an
approximately isotropic distribution \cite{bkv97,bs98}. Such particles
must have a mass $\gtrsim10^{12}\gev$ to account for the highest
energy event, and a lifetime exceeding the age of the universe. A
physically well-motivated candidate with both the required mass and
lifetime is the ``crypton'' --- the analogue of a hadron in the hidden
sector of supersymmetry breaking in string theories
\cite{crypton}.\footnote{Other particle candidates which may be
sufficiently long-lived have also been proposed \cite{other}.} It has
been noted that such particles can readily be produced with a
cosmologically interesting abundance through the time-varying
gravitational field at the end of inflation \cite{gravprod}, or
possibly, during (re)heating after inflation \cite{reheatprod}.

A key test of this model is the expected small anisotropy due to the
offset of the Sun from the centre of the Galaxy
\cite{ddmaniso}.\footnote{If the clustering in the arrival directions
turns out to be statistically significant, it may be an indication of
clumpiness in the halo dark matter distribution, as is indeed expected
for CDM \cite{clustdm}.}  Some authors have argued \cite{fp00,shdmnot}
that present limits on the anisotropy already place severe constraints
on the model; however a recent detailed analysis of the expected
anisotropy signal \cite{efs01} of post-GZK UHECRs shows it to be both
compatible with all present data and moreover capable of being
definitively detected by the Pierre Auger Project which is presently
under construction \cite{auger}, and by the proposed Extreme Universe
Space Observatory \cite{euso}. These experiments will provide a very
substantial increase in the presently meagre statistics of post-GZK
UHECRs and also be able to distinguish between incident photon,
neutrino and nucleon primaries. This is clearly an opportune time to
make predictions of the expected spectrum and composition of UHECRs in
the decaying dark matter (DDM) model.

\section{The decaying dark matter model}
\label{ddm}

The essential idea \cite{bkv97,bs98} here is that because of
gravitational clustering, very massive relic particles would naturally
have a density in the halo of our Galaxy which is enhanced by a factor
of $\sim10^4$ over the cosmic average. This is approximately the same
factor by which the distance to the horizon exceeds the halo radius,
so if these particles decay to produce UHECRs the flux from
cosmologically distant regions will be at best comparable to the flux
from our halo even if there is {\em no} absorption
\cite{exptrev}. Given the GZK attenuation, the extragalactic flux of
protons and photons in this model would be quite negligible compared
to the flux from the Galactic halo, whereas for neutrinos there would
be an extragalactic component comparable to the galactic one
\cite{decay}.

The injection spectrum produced by the decay of a population of
superheavy dark matter particles $X$, with number density $n_X$, is
proportional to the inclusive decay width of $X$:
\begin{equation}
 \label{eq:injection}
 \Phi^\halo (E) = \frac{n_X}{\tau_X} \frac{1}{\Gamma_X}
 \frac{\rmd \Gamma (X \rightarrow h + \dots)}{\rmd E},
\end{equation}
where we have taken the lifetime of $X$ to be longer than the age of
the universe, $\tau_X\gg\,t_0\sim10^{10}$~yr. Photons produced in the
galactic halo can interact with the galactic low frequency radio
background on their way to the Earth; the number of high energy
photons reaching the Earth may then be significantly reduced. However
the radio background is presently very poorly known \cite{radio} so we
do not include photon processing in the halo in this work. We consider
a spherical halo of radius $R_\halo$ and uniform density $n_X^\halo$
(see Ref.\cite{efs01} for a thorough study of UHECR and halo
models). The galactic halo contribution to the UHECR spectrum is then
given by
\begin{equation}
 \label{eq:Flux}
 J^\halo (E) = \frac{1}{4\pi} R_\halo \Phi^\halo (E).
\end{equation}

In order to calculate the inclusive decay width, some assumptions have
to be made about the microphysics of $X$ decay. It is usually assumed
in top-down models (i.e. for both TD and DDM) that the initial decay
is into a parton-antiparton pair. Any angular dependence of the decay
amplitude is immaterial since the flux at Earth is close to
isotropic. Therefore in the following Section (\ref{FFEvolution}) we
shall consider decays into quarks and gluons (and also their
superpartners when allowing for the presence of supersymmetry).

\section{Evolution of fragmentation functions}
\label{FFEvolution}

\subsection{Inclusive Decay Width}

For a particle $X$ with mass $M_X$, decaying into partons which
hadronise into particles of type $h$ (carrying a fraction $x$ of the
maximum available momentum $M_X/2$, and a fraction $z$ of the parton
momentum), the inclusive decay width can be factorised as \cite{ESW}
\begin{equation} 
 \label{eq:DecayRate}
 \frac{1}{\Gamma_X}\frac{\rmd \Gamma(X \rightarrow h + \dots)}{\rmd x} = 
  \sum_a \int^1_x \frac{\rmd z}{z}\, \frac{1}{\Gamma_a}
  \left.\frac{\rmd \Gamma_a(y,\mu^2,M_X^2)}{\rmd y}\right|_{y=x/z} 
  D^h_a (z,\mu^2).
\end{equation}
The first factor, the decay width of $X$ into parton $a$,
$\rmd\Gamma_a/\rmd\,y$, is calculable in perturbation theory; in
lowest order and for 2-body decay it is proportional to
$\delta(1-y)$. The second factor, the non-perturbative $D^h_a$, is the
fragmentation function (FF) for particles of type $h$ from partons of
type $a$. Both factors depend on the factorisation scale $\mu$. The
dependence on $\mu$ of the physical inclusive decay width would cancel
out if the calculation could be carried out to all orders in
perturbation theory but in a finite-order calculation some dependence
on $\mu$ remains to orders higher than those calculated. The choice
$\mu\sim\,M_X$ is the most appropriate \cite{NasonWebber}. Every
perturbative term in the above factorisation depends also on an
arbitrary renormalization scale; following the usual convention, we
take the renormalization scale to equal the factorisation scale.

The cancellation of the dependence on $\mu$ between the perturbative
factors and the FFs constrains the latter to satisfy the
Dokshitzer--\-Gribov--\-Lipatov--\-Altarelli--\-Parisi (DGLAP)
equations \cite{AltarelliParisi,DGL}. Given experimental data on
hadronisation at some low energy scale, say $M_Z$, an initial set of
fragmentation functions $D^h_a(x,M_Z^2)$ can be extracted. These can
then be evolved using the DGLAP equations to obtain the FFs at some
higher scale $D^h_a(x,M_X^2)$. This evolution gives rise to scaling
violations in the $x$--dependence of the inclusive decay width of $X$
in exactly the same way that scale-dependence of structure functions
and FFs produce scaling violations in experimentally measured hadronic
cross-sections \cite{ESW}.

The DGLAP equations can be written as
\begin{equation}
 \label{eq:AP}
 \frac{\partial D^h_a(x,\mu^2)}{\partial\ln\mu^2} = 
 \sum_b \frac{\alpha_s(\mu^2)}{2\pi} 
 P_{ba}(x,\alpha_s(\mu^2)) \otimes D^h_b(x,\mu^2),
\end{equation}
where $\alpha_s(\mu^2)$ is the strong coupling constant and
$P_{ba}(x,\alpha_s)$ is the splitting function for the parton
branching $a\rightarrow\,b$. Here the convolution of two functions
$A(x)$ and $B(x)$ is defined as
\begin{equation}
 \label{eq:Convolution}
 A(x) \otimes B(x) \equiv \int^1_x \frac{\rmd z}{z}\, A(z) B(\case{x}{z}).
\end{equation}
The splitting functions can be expanded perturbatively:
\begin{equation}
 \label{eq:LO}
 P_{ba}(x,\alpha_s) = P_{ba}(x) + {\cal O}(\alpha_s).
\end{equation}
We limit our study to leading order in $\alpha_s$ and therefore ignore
${\cal O}(\alpha_s)$ corrections to the splitting functions. It is
also convenient to define the following dimensionless evolution
parameter
\begin{equation}
 \label{eq:Tau}
 \tau \equiv \frac{1}{2\pi b}\ln\frac{\alpha_s(\mu^2_0)}{\alpha_s(\mu^2)}\ ,
\end{equation}
$b$ being the coefficient in the leading order $\beta$-function
governing the running of the strong coupling:
$\beta(\alpha_s)=-b\alpha_s^2$. We take $D^h_a$ to represent the
sum of particle $h$ and, if different, its antiparticle $\bar{h}$.

\subsection{Standard Model Equations}

The Standard Model DGLAP equations for the evolution of fragmentation
functions are well-known \cite{ESW,NasonWebber}. There are two parton
species: quarks $q_k$, $k=1,\dots n_\F$, and gluons $g$, with $n_\F$
the total number of flavours. Conventionally one defines the following
linear combinations (for ease of notation the superscript $h$ is
omitted):
\begin{eqnarray}
 \label{eq:Comb1}
 D_{q^+_k} &\equiv& D_{q_k}+ D_{\bar{q}_k}, \\
 \label{eq:Comb2}
 D_q &\equiv& \sum_k D_{q^+_k}, \\
 \label{eq:Comb3}
 D_{q^-_k} &\equiv& D_{q_k}- D_{\bar{q}_k}, \\
 \label{eq:Comb4}
 D_{Q_k} &\equiv& D_{q^+_k}- \frac{1}{n_\F}D_q.
\end{eqnarray}
The non-singlet functions $D_{q^-_k}$ and $D_{Q_k}$ obey the
equations
\begin{eqnarray}
 \label{eq:NonSinglet1}
 \partial_\tau D_{q^-_k} &=& P_{qq} \otimes  D_{q^-_k}, \\
 \label{eq:NonSinglet2}
 \partial_\tau D_{Q_k} &=& P_{qq} \otimes  D_{Q_k},
\end{eqnarray}
while the evolution of the singlet function $D_q$ is coupled to that
of the gluon function $D_g$ as
\begin{equation}
  \label{eq:SM2X2}
  \partial_\tau 
  \left(
    \begin{array}{l}
      D_q \\
      D_g
    \end{array}
  \right)
  =
  \left(
    \begin{array}{cc}
      P_{qq} & 2n_{\F}P_{gq} \\
      P_{qg} & P_{gg}
    \end{array}
  \right)
  \otimes
  \left(
    \begin{array}{l}
      D_q \\
      D_g
    \end{array}
  \right).
\end{equation}
The splitting functions were calculated in
Refs.\cite{AltarelliParisi,ESW}. Given the FFs at some initial scale
$\mu_0$ for the quarks $q_k$ and gluon $g$, we can now determine their
evolved values at some other scale $\mu$ to leading order in
$\alpha_s$, using Eqs.~(\ref{eq:NonSinglet1}--\ref{eq:SM2X2}).

\subsection{Supersymmetric Equations}

In a supersymmetric (SUSY) model, besides the quarks and gluons one
has their superpartners: squarks $s_k$ and gluinos $\lambda$. In
addition to the linear combinations~(\ref{eq:Comb1}--\ref{eq:Comb4})
we now define
\begin{eqnarray}
  \label{eq:SCombinations}
  D_{s^+_k} &\equiv& D_{s_k}+ D_{\bar{s}_k}, \\
  D_s &\equiv& \sum_k D_{s^+_k}, \\
  D_{s^-_k} &\equiv& D_{s_k}- D_{\bar{s}_k}, \\
  D_{S_k} &\equiv& D_{s^+_k}- \frac{1}{n_\F}D_s.
\end{eqnarray}
The non-singlet function $ D_{q^-_k}$ and $ D_{s^-_k}$ evolve together, 
as do $D_{Q_k}$ and $D_{S_k}$:
\begin{eqnarray}
  \label{eq:SUSY2X2a}
  \partial_\tau 
  \left(
    \begin{array}{l}
     D_{q^-_k} \\
     D_{s^-_k}
    \end{array}
  \right)
  &=&
  \left(
    \begin{array}{cc}
      P_{qq} & P_{sq} \\
      P_{qs} & P_{ss}
    \end{array}
  \right)
  \otimes
  \left(
    \begin{array}{l}
      D_{q^-_k} \\
      D_{s^-_k}
    \end{array}
  \right), 
  \\
  \label{eq:SUSY2X2b}
  \partial_\tau 
  \left(
    \begin{array}{l}
      D_{Q_k} \\
      D_{S_k}
    \end{array}
  \right)
  &=&
  \left(
    \begin{array}{cc}
      P_{qq} & P_{sq} \\
      P_{qs} & P_{ss}
    \end{array}
  \right)
  \otimes
  \left(
    \begin{array}{l}
      D_{Q_k} \\
      D_{S_k}
    \end{array}
  \right).
\end{eqnarray}
The singlet functions for quarks and squarks, $D_q$ and $D_s$, are
coupled to the gluon and gluino functions, $D_g$ and $D_\lambda$, as
\begin{equation}
  \label{eq:SUSY4X4}
 \partial_\tau 
  \left(
    \begin{array}{l}
      D_q \\
      D_g \\
      D_s \\
      D_\lambda
    \end{array}
  \right)
  =
  \left(
    \begin{array}{cccc}
      P_{qq} & 2n_{\F}P_{gq} & P_{sq} & 2n_{\F}P_{\lambda q} \\
      P_{qg} & P_{gg} & P_{sg} & P_{\lambda g}  \\
    P_{qs} & 2n_{\F}P_{gs} & P_{ss} & 2n_{\F}P_{\lambda s} \\
    P_{q\lambda} & P_{g\lambda} & P_{s\lambda} & P_{\lambda\lambda}
    \end{array}
  \right)
  \otimes
  \left(
    \begin{array}{l}
      D_q \\
      D_g \\
      D_s \\
      D_\lambda
    \end{array}
  \right). 
\end{equation}
In leading order, Eqs.~(\ref{eq:SUSY2X2a}--\ref{eq:SUSY4X4}) allow us
to calculate the fragmentation functions for all quark and squark
flavours, gluons and gluinos at some scale $\mu$, given their values
at some initial scale $\mu_0$.

Incidentally, using the following relationships among the SUSY
splitting functions \cite{KounnasRoss},
\begin{eqnarray}
  \label{eq:SUSYrelations}
  P_{qq}+P_{sq} &=& P_{ss}+P_{qs}, \\
  P_{gq}+P_{\lambda q} &=& P_{\lambda s}+P_{gs}, \\
  P_{qg}+P_{sg} &=& P_{s\lambda}+P_{q\lambda}, \\
  P_{gg}+P_{\lambda g} &=& P_{\lambda \lambda}+P_{g\lambda}, 
\end{eqnarray}
one can reduce the $4\times 4$ matrix equation (\ref{eq:SUSY4X4}) to the
$2\times2$ matrix equation
\begin{equation}
  \label{eq:SUSY2X2}
  \partial_\tau 
  \left(
    \begin{array}{l}
      D_q - D_s     \\
      D_g - D_\lambda
    \end{array}
  \right)
  =
  \left(
    \begin{array}{cc}
      P_{qq}-P_{qs} & 2n_{\F}(P_{gq}-P_{gs}) \\
      P_{qg}-P_{q\lambda} & P_{gg}-P_{g\lambda}
    \end{array}
  \right)
  \otimes
  \left(
    \begin{array}{l}
      D_q - D_s     \\
      D_g - D_\lambda
    \end{array}
  \right).
\end{equation}
The SUSY DGLAP equations have been given in the literature to leading
order for structure functions \cite{KounnasRoss,JonesLlewellyn}. Here
we have presented their form for FFs. It is easy to see that (except
for Eq.~(\ref{eq:SUSY2X2})) one just needs to transpose the matrix
elements, keeping the $n_\F$ factors in the same place, to move from
structure function to fragmentation function equations. For
Eq.~(\ref{eq:SUSY2X2}), the relative minus sign between $q,s$ and
$g,\lambda$ changes to a plus sign and there is a rearrangement of the
splitting functions in the $2\times 2$ matrix. The SUSY splitting
functions were calculated in Refs.\cite{KounnasRoss,JonesLlewellyn}.

\subsection{Numerical Evolution: Algorithm and Initial Conditions}

Several numerical algorithms and their implementation codes are
available to solve the set of integro-differential DGLAP equations for
structure functions when quarks and gluons are the only particles
inside the hadrons \cite{FurmanskiPetronzio,SMcodes}. The
supersymmetric evolution of structure functions has been considered in
Ref.\cite{Coriano}. These codes are suitable for collider physics but
are rather unwieldy to use for studying cosmic ray production. We have
generalised the Laguerre method \cite{FurmanskiPetronzio} to include
SUSY evolution for fragmentation functions and written a numerical
code to calculate cosmic ray production from superheavy particle
decay. The main idea of the Laguerre method is to expand
Eq.~(\ref{eq:AP}) in Laguerre polynomials; the coefficients of the
series expansion are given by simple recursive relations appropriate
for computer calculations. These recursive relations have been given
so far for the 1-dimensional case
(Eqs.~\ref{eq:NonSinglet1}--\ref{eq:NonSinglet2}) and the
2-dimensional case (Eq.~\ref{eq:SM2X2}). However to study the general
SUSY evolution requires one to work out the recursive relations for
the 4-dimensional case (Eq.~\ref{eq:SUSY4X4}). This is not a trivial
step if one follows the procedure given for lower dimensional cases in
the literature. Hence we have calculated the general recursive
relations for any dimension. The details of this procedure, as well as
our numerical code, will be given elsewhere \cite{Toldra}.

\begin{figure}[htb]
  \epsfxsize\hsize\epsffile{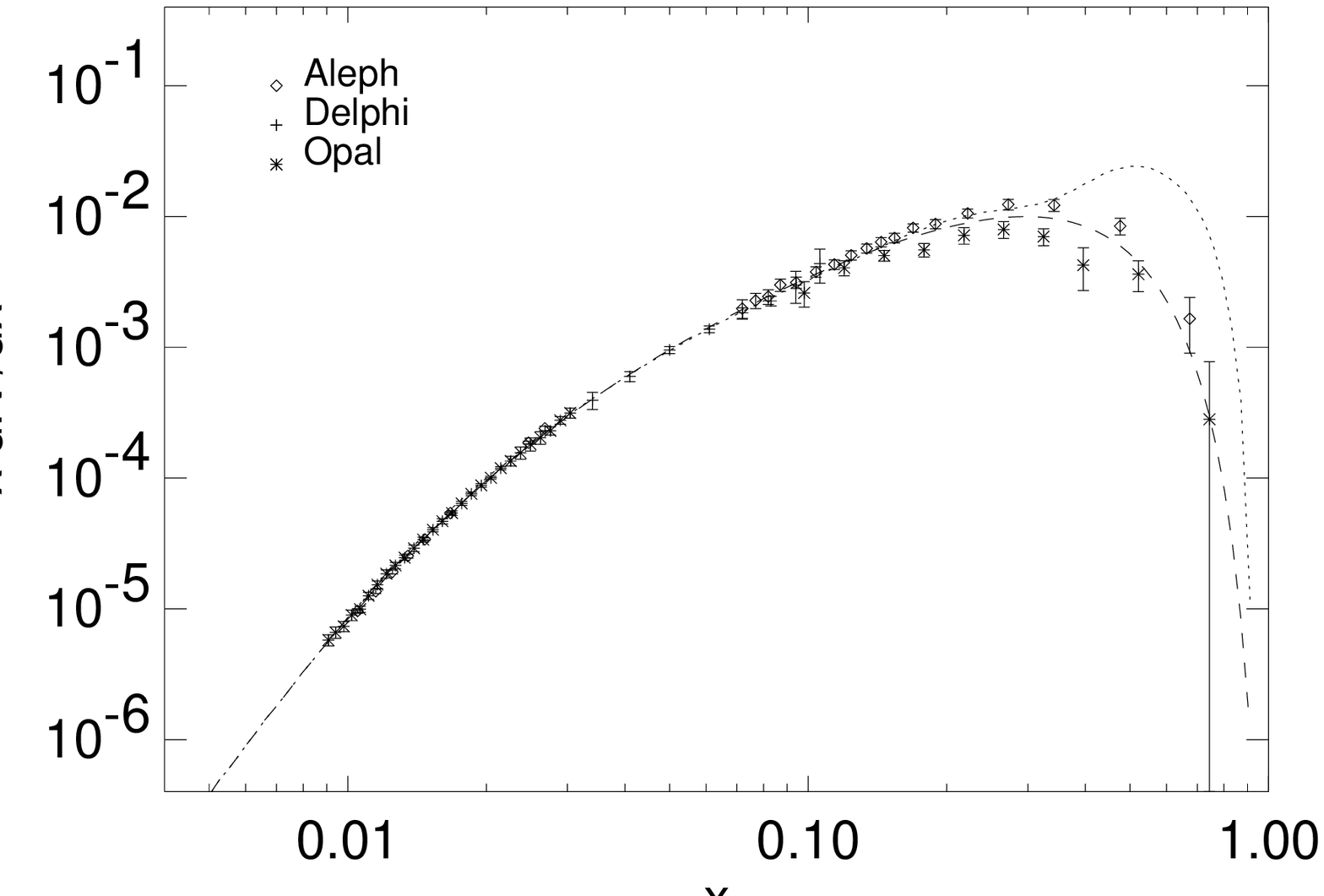}
  \bigskip\bigskip
  \caption{The fragmentation spectrum of protons at the $Z^0$ mass
  peak, as measured by ALEPH, DELPHI and OPAL at LEP; the dashed line is a
  parametric fit \protect\cite{Rubin}. The dotted line is the spectrum
  simulated with HERWIG, illustrating its tendency to overproduce
  baryons at high $x$.}
  \label{fig:LEP_p} 
\end{figure}

\begin{figure}[htb]
  \epsfxsize\hsize\epsffile{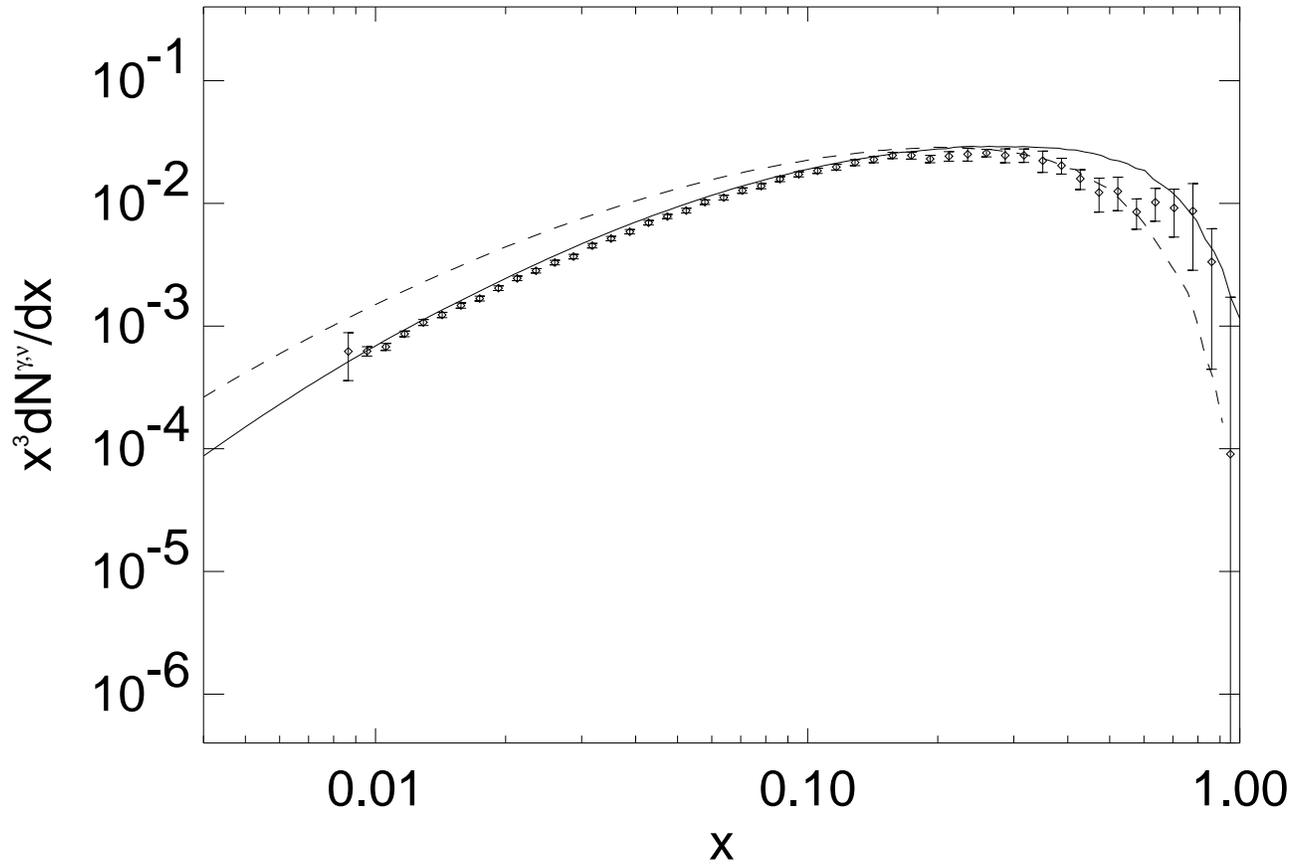}
  \bigskip\bigskip
  \caption{The fragmentation spectrum of photons at the $Z^0$ mass
  peak, as measured by ALEPH at LEP; the solid line is the spectrum
  simulated with HERWIG. The fragmentation spectrum of neutrinos from
  HERWIG is also shown as the dashed line.}
  \label{fig:LEP_gamnu} 
\end{figure}

We begin with FFs at the energy scale $\mu_0=M_Z$ and evolve them to
the final energy scale $\mu=M_X$ using Eq.~(\ref{eq:AP}). For the
initial fragmentation function of baryons,
$D^p_a(x,M_Z^2)+D^n_a(x,M_Z^2)$, we adopt the fit performed in
Ref.~\cite{Rubin} to LEP hadronic data \cite{LEP}, as shown in
Fig.~\ref{fig:LEP_p}. For photons and neutrinos we generate initial
data at the $Z$ peak using the QCD Monte Carlo event generator HERWIG
\cite{HERWIG}. Comparison with LEP data shows that although HERWIG
overproduces baryons at high $x$ \cite{Rubin,kupco}, its photon and
meson output at the $Z$ peak matches the experimental spectra
remarkably well. Since neutrinos mainly come from charged pion and
kaon decays, one can thus be confident in taking the HERWIG generated
FF for neutrinos as the initial condition for the evolution. There is
also a sizable contribution from heavy flavour decays to the neutrino
spectrum at high $x$ which is explicitly taken into account by
HERWIG. Figure~\ref{fig:LEP_gamnu} shows our initial spectra at the
energy scale $M_Z$ for photons and neutrinos together with LEP data
for photons.

We assume flavour universality in the decay of $X$, hence we consider
only the coupled singlet quark and gluon evolution equations
(\ref{eq:SM2X2}) for the Standard Model (SM), and the coupled singlet
quark, gluon, singlet squark and gluino evolution equations
(\ref{eq:SUSY4X4}) for a SUSY model. A (s)parton is not included in
the evolution as long as the energy scale is lower than its mass; when
the threshold for its production is crossed, it is added to the
evolution equations with an initially {\em vanishing} FF and it is
assumed to be a relativistic particle.

In the SM case we evolve the $q$ and $g$ initial fragmentation
functions from $M_Z$ to $M_t$, the top quark mass, with the number of
flavours set to $n_\F=5$, and then evolve from $M_t$ to $M_X$ with
$n_\F=6$. (However assuming $n_\F=6$ in the whole range from $M_Z$ to
$M_X$ does not introduce any significant difference in the final
spectrum.)

In the SUSY case we evolve the $q$ and $g$ initial fragmentation
functions from $M_Z$ to the supersymmetry breaking scale $M_\SUSY>M_t$
using the SM equations to obtain $D^h_i(x,M^2_\SUSY)$, with $i=q,g$.
Then we take $D^h_i(x,M^2_\SUSY)$, $i=q,g$, and
$D^h_j(x,M^2_\SUSY)=0$, $j=s,\lambda$, and evolve them from $M_\SUSY$
to $M_X$ using the SUSY equations. All spartons are taken to be
degenerate with a common mass $M_\SUSY$. (In the context of structure
functions, a SUSY model with different masses for $s$ and $\lambda$ was
studied~\cite{Coriano}, finding no significant difference with models
having the same mass for all superpartners. One might expect the same
result to hold for FFs.)

For very small $x$ values, $x\lesssim0.001$, coherent gluon emission
becomes important. In this regime Eq.~(\ref{eq:AP}) no longer holds
and the kernel of the integral must be modified as
$D_a(x/z,\mu^2)\rightarrow D_a(x/z,z^2\mu^2)$. This is the Modified
Leading Logarithm Approximation (MLLA) \cite{ESW,MLLA} which has been
used by other authors to calculate the cosmic ray spectrum in top-down
models \cite{bbv98,bkv97,bk98}. However the MLLA spectrum {\em cannot}
be normalised since a very small (and uncertain) fraction of the total
energy is released in this kinematic region; moreover in the context
of the DDM model for a particle mass of order the hidden-sector scale,
the observed highest energy cosmic rays correspond to large values of
$x$ \cite{bs98,Sarkar}. There are also other corrections to the DGLAP
equations arising from a more rigorous treatment of threshold effects
\cite{NasonWebber} and, of course, next-to-leading order corrections
to $\alpha_s$ which we have not considered. Nevertheless, the level of
precision afforded by the DGLAP equations in leading order, when
calculating scaling violations in the inclusive decay of $X$, is quite
adequate for comparison with the present experimental data on UHECRs.

For $x\sim1$ there are large uncertainties (see Fig.~\ref{fig:LEP_p})
in the experimental determination of the FFs \cite{LEP}. Generating
values for these using HERWIG requires rather large amounts of
computational time since very few particles fall in the bins with
$x>0.8$, if the number of simulated events is kept to a reasonable
number \cite{bs98}. Furthermore, the algorithms which solve the DGLAP
equations show poor convergence in this regime. These problems
motivate an alternative approach to calculating high $x$
fragmentation. One can take a power-law fit to the experimental data
at $M_Z$, and then use the Mellin transform technique \cite{ESW} to
calculate its evolved value at $M_X$. If at $\tau_0$ one has the fit
$D_a(x,\tau_0)=K (1-x)^k$, where $K$ and $k$ are constants, then when
$x\rightarrow1$ we find
\begin{equation}
  \label{eq:Mellin}
  D_a(x,\tau) = K\left[\frac{\alpha_s(\tau_0)}{\alpha_s(\tau)}\right]^{B_a}
  \frac{\Gamma(k+1)}{\Gamma(A_a(\tau-\tau_0)+k+1)}
  (1-x)^{k+A_a(\tau-\tau_0)},
\end{equation}
where in the SM
\begin{eqnarray}
 A_q = 2C_\F, \quad && B_q = 
  \case{C_\F}{2\pi b}\left(\case{3}{2}-2\gamma\right), \\
 A_g = 2C_\A, \quad && B_g = \case{1}{2\pi b}
  \left[C_\A\left(\case{11}{6}-2\gamma\right)-\case{2}{3}T_\R n_\F\right],
\end{eqnarray}
and in a SUSY model
\begin{eqnarray}
 A_q = A_s = 2C_\F, \quad && B_q = B_s = \case{C_\F}{2\pi b}(1-2\gamma), \\
 A_g = A_\lambda = 2C_\A, \quad &&
 B_g = B_\lambda = \case{1}{2\pi b}
  \left[C_\A\left(\case{3}{2}-2\gamma\right)-T_\R n_\F\right].
\end{eqnarray}
The $SU(3)$ fundamental and adjoint group factors are respectively
$C_\F=4/3$, $C_\A=3$, $b$ is the leading order coefficient of the
$\beta$-function (\ref{eq:Tau}), $T_\R=1/2$, and $\gamma=0.57721\dots$
is the Euler constant.  Eq.~(\ref{eq:Mellin}) shows that for
$x\rightarrow\,1$, a power-law $D_a$ evolves at higher energy into a
power-law with a bigger power index: the higher the energy the faster
the drop in $(1-x)$.

\subsection{Numerical Evolution: Results and Discussion}

\begin{figure}[htb]
  \epsfxsize\hsize\epsffile{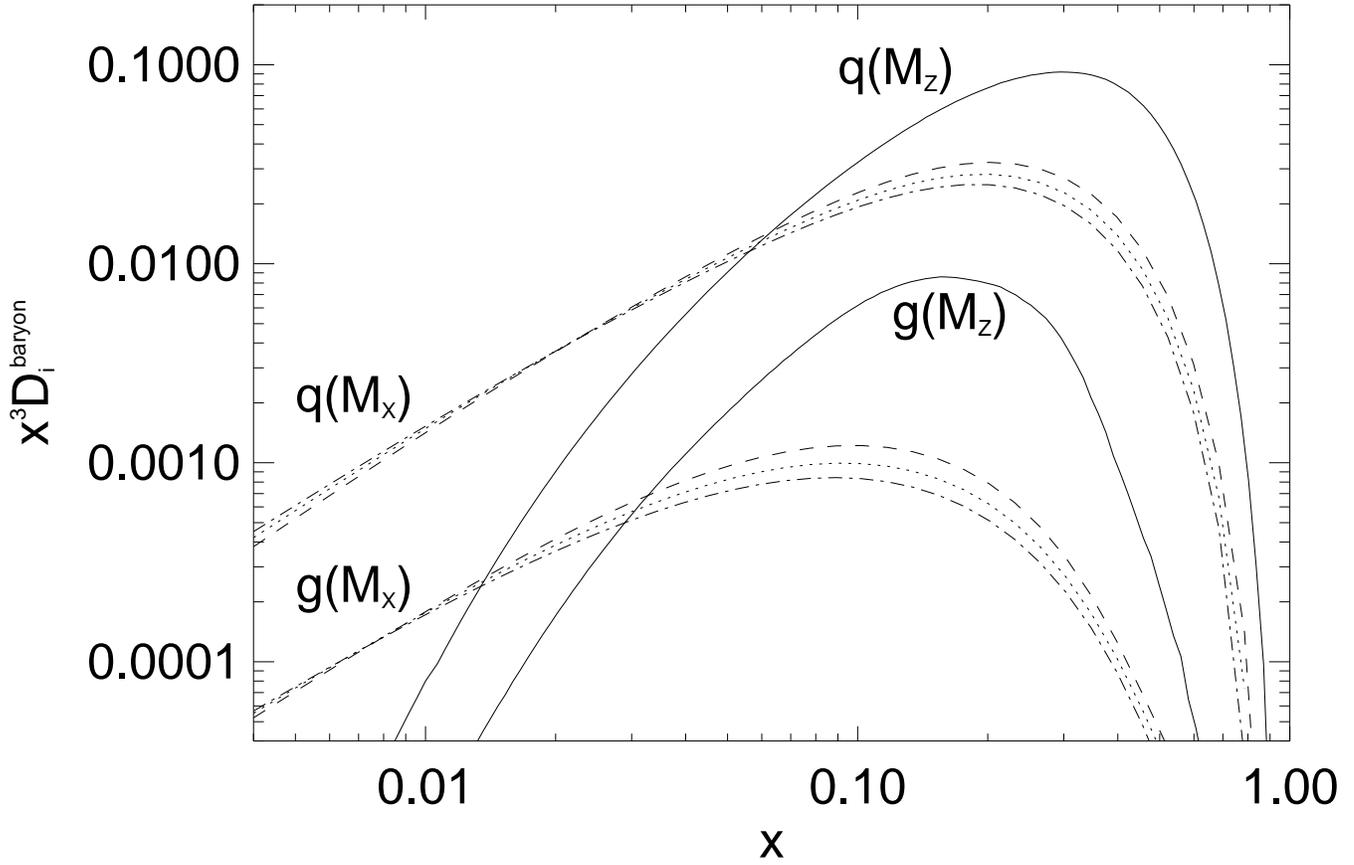}
  \bigskip\bigskip
  \caption{Standard Model fragmentation functions for baryons from
  quarks and gluons, at the initial scale $M_Z$ (solid lines) and
  evolved to a decaying particle mass scale of $10^{10}\gev$ (dashed
  line), $10^{12}\gev$ (dotted line) and $10^{14}\gev$ (dot-dashed
  line), illustrating scaling violations}  
\label{fig:qgSM} 
\end{figure}

First we present the evolution of the baryon fragmentation function in
the SM. In Fig.~\ref{fig:qgSM} we show the fragmentation functions for
baryons ($p$ and $n$) from quarks and gluons at the scale $M_Z$ as
measured at LEP, and their evolved shape at $M_X=10^{10},10^{12}$ and
$10^{14}$~GeV. (Following the standard convention we always plot the
quantity $x^3D_a(x,\mu^2)$, which is proportional to the cosmic ray
energy spectrum $E^3J(E)$.) As the final scale increases, the number
of particles grows at low $x$ and decreases at high $x$ as is
well-known from many previous studies of scaling violations.

\begin{figure}[htb]
  \epsfxsize\hsize\epsffile{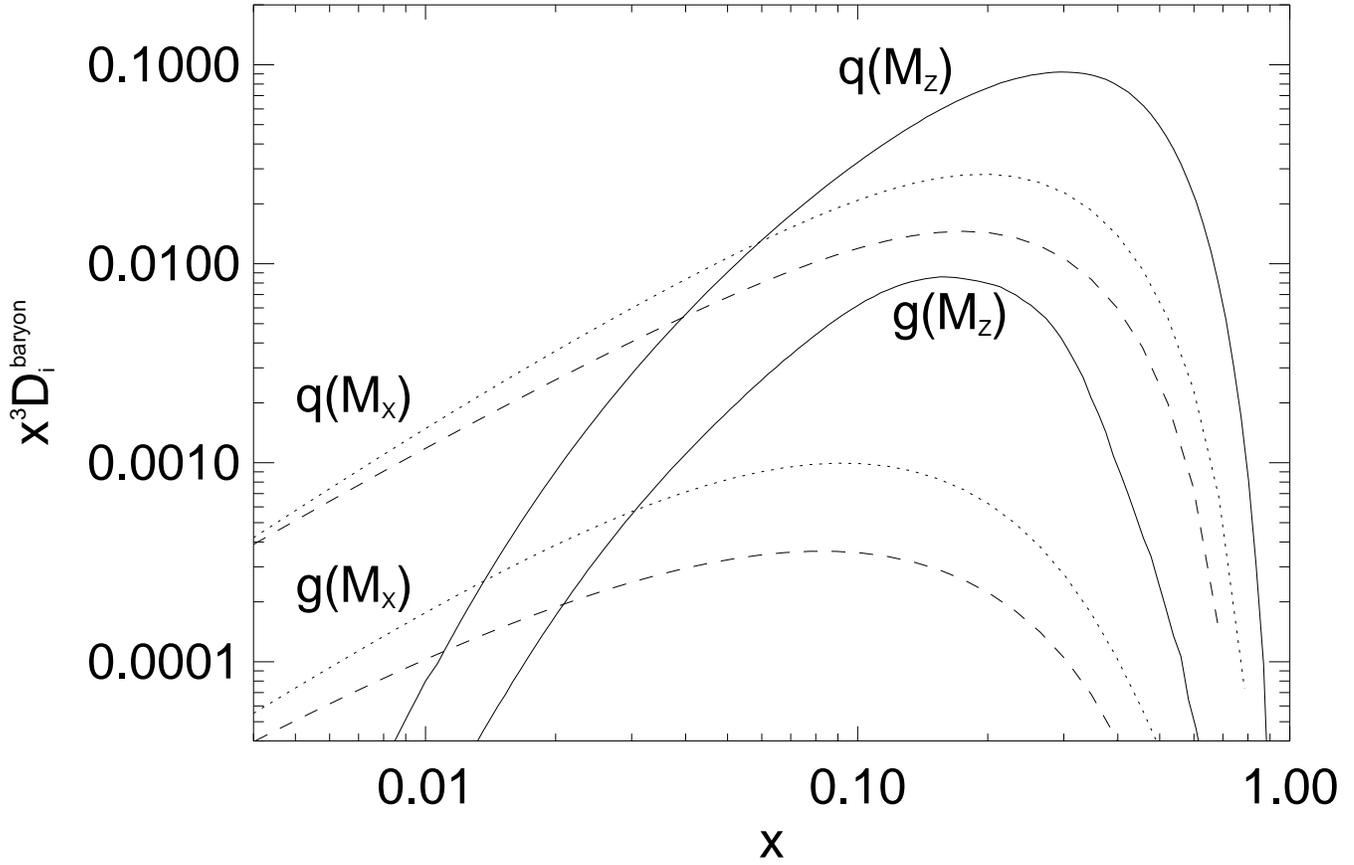} 
  \bigskip\bigskip
  \caption{Fragmentation functions for baryons from
  quarks and gluons, at the initial scale $M_Z$ (solid lines) and
  evolved to a decaying particle mass scale of $10^{12}\gev$, for
  SM evolution (dotted lines), and, the more pronounced, SUSY 
  evolution (dashed lines) taking $M_\SUSY=400\,\gev$.}
  \label{fig:qgSM_SUSY}
\end{figure}

Next we compare the SUSY evolution of FFs with SM evolution. In
Fig.~\ref{fig:qgSM_SUSY} we show the common initial baryon FF at
$M_Z$, its shape after SM evolution up to $M_X=10^{12}$~GeV, and its
evolved shape at the same final scale after SUSY has been switched on
at $M_\SUSY=400$~GeV. It is clear that the SUSY curve has evolved {\em
further} than the SM curve, chiefly due to the different running of
$\alpha_s(\mu^2)$. In a SUSY model $\alpha_s$ decreases with
increasing energy more slowly than in the SM because of the increased
contribution to the $\beta$-function from the SUSY partners. Since the
rate of change $\partial_{\ln \mu^2} D_a(x,\mu^2)$ is proportional to
$\alpha_s$ (see Eq.~\ref{eq:AP}), a bigger $\alpha_s$ translates into
a larger amount of evolution. In other words, given the same initial
and final scales we obtain $\tau_{\SUSY}(M_X)>\tau_{\SM}(M_X)$, using
Eq.~(\ref{eq:Tau}).

\begin{figure}[htb]
  \epsfxsize\hsize\epsffile{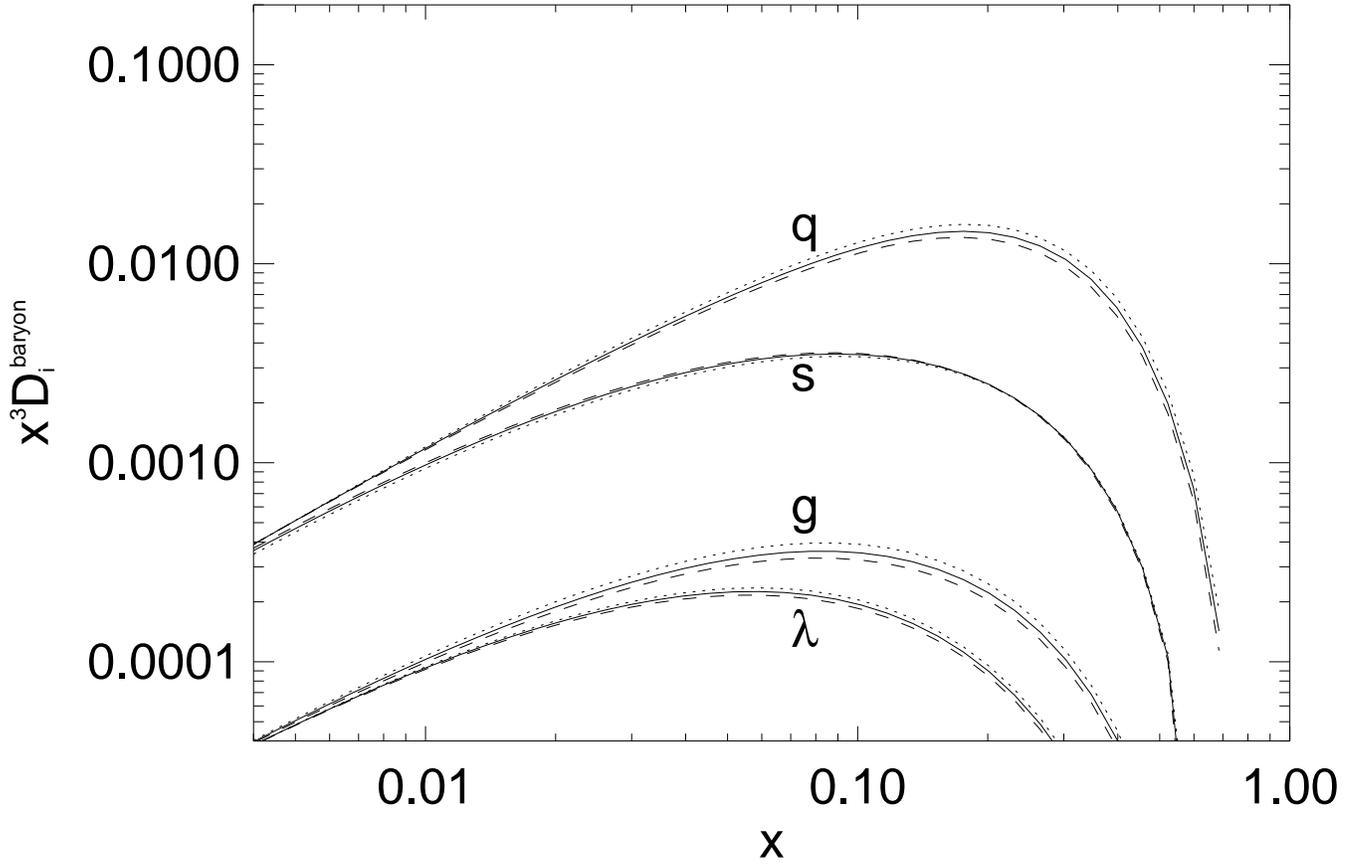}
  \bigskip\bigskip
  \caption{Dependence of (s)parton fragmentation functions evolved
  from $M_Z$ up to $M_X=10^{12}\gev$ on $M_\SUSY$ = $200\gev$ (dashed
  lines), $400\gev$ (solid lines), $1$~TeV (dotted lines).}
  \label{fig:qgslSUSY}
\end{figure}

Figure~\ref{fig:qgslSUSY} shows the quark and gluon FFs at $M_Z$,
their evolved values at $M_X=10^{12}$~GeV using the SUSY equations for
scales larger than $M_\SUSY$, and the radiatively generated squark and
gluino functions, all at the same final scale $M_X$. We find that
starting from vanishing values at $M_\SUSY$ the squark and gluino
functions start to grow and catch up rapidly with the quark and gluon
functions, respectively, at small $x$. This behaviour can be
understood qualitatively if one bears in mind that at low $x$ the
leading splitting function for quarks is
$2n_{\F}P_{gq}\sim4n_{\F}C_{\F}/x$, which is equal to the leading
splitting function for squarks $2n_{\F}P_{gs}\sim4n_{\F}
C_{\F}/x$. For gluons and gluinos the leading splitting functions tend
as well to a common value, $P_{gg}\sim2C_\A/x$ and
$P_{g\lambda}\sim2C_\A/x$, which is however different from that of
quarks and squarks.

SUSY evolution does not depend strongly on the chosen supersymmetry
breaking scale $M_\SUSY$. In Fig.~\ref{fig:qgslSUSY} we show the
curves obtained taking $M_\SUSY=200,400$~GeV and 1~TeV. The higher the
value of $M_\SUSY$, the less evolved the final curves for $q$ and
$g$. This follows from our earlier comparison of SM vs. SUSY
evolution. If SUSY switches on later (higher $M_\SUSY$) the energy
range over which the SM equations hold is larger. As we have seen
already, DGLAP evolution is slower when just the SM equations are
employed.

\begin{figure}[htb]
 \epsfxsize15cm\epsffile{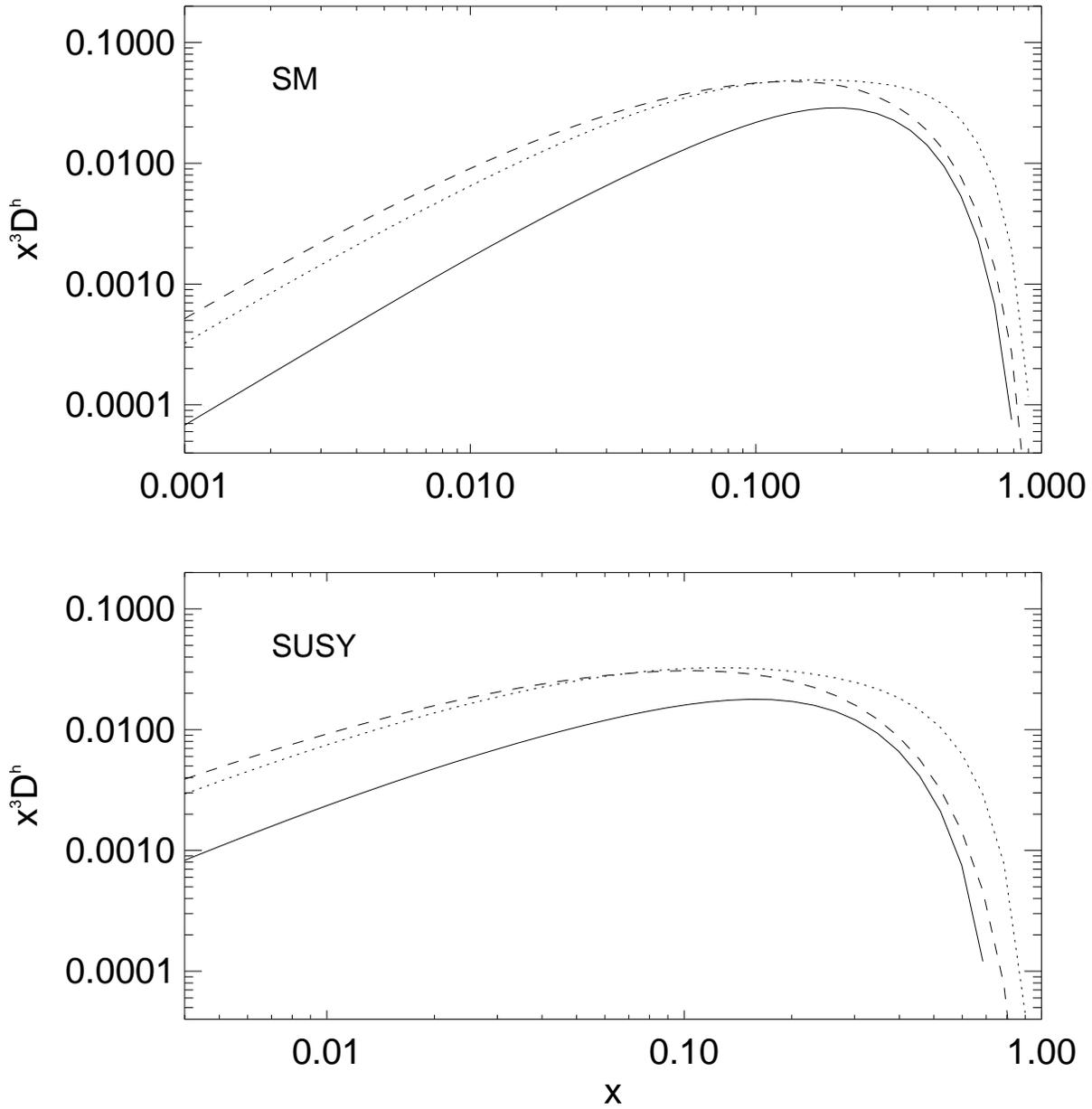} 
  \bigskip\bigskip
  \caption{Fragmentation functions for baryons (solid lines), photons
  (dotted lines) and neutrinos (dashed lines) evolved from $M_Z$ up to
  $M_X=10^{12}\gev$ for the SM (top panel) and for SUSY with
  $M_\SUSY=400\gev$ (bottom panel).}
  \label{fig:allPartSM-SUSY}
\end{figure}

The evolution of the photon and neutrino spectra can also be studied
using the DGLAP equations. We take HERWIG simulated data for the
initial FFs for photons and neutrinos (sum over all three flavours) at
$M_Z$ and use the SM and SUSY evolution equations (\ref{eq:SM2X2})
and~(\ref{eq:SUSY4X4}) to obtain their spectra at $M_X$. We plot in
Fig.~\ref{fig:allPartSM-SUSY} the fragmentation functions for baryons,
photons and neutrinos for a decaying particle of mass
$M_X=10^{12}$~GeV, in both the SM and in a SUSY model with
$M_\SUSY=400$~GeV. As pointed out earlier we assume flavour symmetry
in the decay of $X$; moreover all the calculated FFs are
colour-weighted. Hence for the SM the total contribution to the FF for
particle of type $h$ is just the sum of quark singlet and gluon FFs
\begin{equation}
  \label{eq:colourSumSM}
  D^h \equiv D^h_q + D^h_g,
\end{equation}
while in a SUSY model we take the sum of quark and squark singlets,
gluon and gluino FFs
\begin{equation}
  \label{eq:colourSumSUSY}
  D^h \equiv D^h_q + D^h_g + D^h_s + D^h_\lambda.
\end{equation}

\section{Ultra High Energy Cosmic Ray Spectrum}

We can now translate the calculated fragmentation functions into the
expected cosmic ray spectrum in order to confront the observational
data. In the previous Section (\ref{FFEvolution}) we have calculated
quark singlet and gluon functions for the SM, and quark singlet,
gluon, squark singlet and gluino functions for SUSY. In the absence of
a specific model for the different branching ratios we have weighted
all the (s)parton contributions evenly but for colour weights as in
Eqs.~(\ref{eq:colourSumSM}--\ref{eq:colourSumSUSY}). Thus the quark
contribution is the most important as seen from Figs. \ref{fig:qgSM}
and \ref{fig:qgslSUSY}; in the SUSY case the squark contribution is
also relevant for $x\le 0.01$. For 2-body decay, $x=2E/M_X$, so from
Eqs. (\ref{eq:injection}--\ref{eq:DecayRate}) we have the following
expression for the flux of particle $h$:
\begin{equation}
  \label{eq:E3Flux}
  E^3 J^\halo (E) = B x^3 D^h (x, M^2_X).
\end{equation}
We have multiplied the flux by $E^3$, as is usual, to emphasise the
structure in the spectrum near the GZK energy. The normalisation
factor $B$ is common for the galactic halo flux of baryons, neutrinos
and photons, and determines the quantity $n_X/\tau_X$ (see
Eq.~\ref{eq:DecayRate}).

Let us now compare the calculated cosmic ray flux to the published
data from Fly's Eye \cite{fe}, AGASA \cite{agasa}, Haverah Park
\cite{hp} and Yakutsk \cite{yak}. In Ref.~\cite{NaganoWatson} these
data have been carefully assessed for mutual consistency and
appropriate adjustments made to the energy calibration, e.g. the
original AGASA data \cite{agasa} was reduced in energy by 10\% to
match the Akeno 1-km$^2$ array data which covers the better explored
energy region $\sim10^{15}-10^{18}\ev$ \cite{akeno}. The authors
recommend adoption of the following standard differential energy
spectrum {\em below} the GZK energy, in the range
$4\times10^{17}\ev<E<6.3\times10^{18}\ev$:
\begin{equation}
 \label{eq:LowComp}
  J(E) = (9.23\pm0.65)\times10^{-33}\;
  \mbox{m}^{-2}\mbox{s}^{-1}\mbox{sr}^{-1}\mbox{eV}^{-1}
  \left(\frac{E}{6.3\times10^{18}\;\mbox{eV}}\right)^{-3.20\pm0.05},
\end{equation}
with the spectrum flattening at higher energies as
$J(E)\propto\,E^{-2.75\pm0.2}$ upto the GZK energy, and extending
further to at least $3\times10^{20}\ev$ \cite{NaganoWatson}. Thus the
UHECR spectrum can naturally be interpreted \cite{fe} as the
superposition of the `low energy' component (\ref{eq:LowComp}), and
the new `flat' component that extends into the post-GZK region. The
former is presumably galactic in origin (consistent with the detection
of anisotropy at $\sim10^{18}\ev$ \cite{anisofe,diskaniso}), while the
latter is interpreted \cite{bs98} as produced by the decay of a
superheavy particle population in the galactic halo. Taking baryons to
be the dominant primary UHECRs as indicated by experiment, the total
flux is
\begin{equation}
  \label{eq:TotalFlux}
  E^3 J(E) = \frac{k}{E^m} + B x^3 D^\baryon (x,M^2_X),
\end{equation}
where the values of $k$ and $m$ can be read off
Eq.~(\ref{eq:LowComp}). Note that since $D^\baryon$ and $D^\gamma$
have a similar shape, taking photons to be the primaries would just
alter the normalisation $B$.

\begin{figure}[htb]
  \epsfxsize\hsize\epsffile{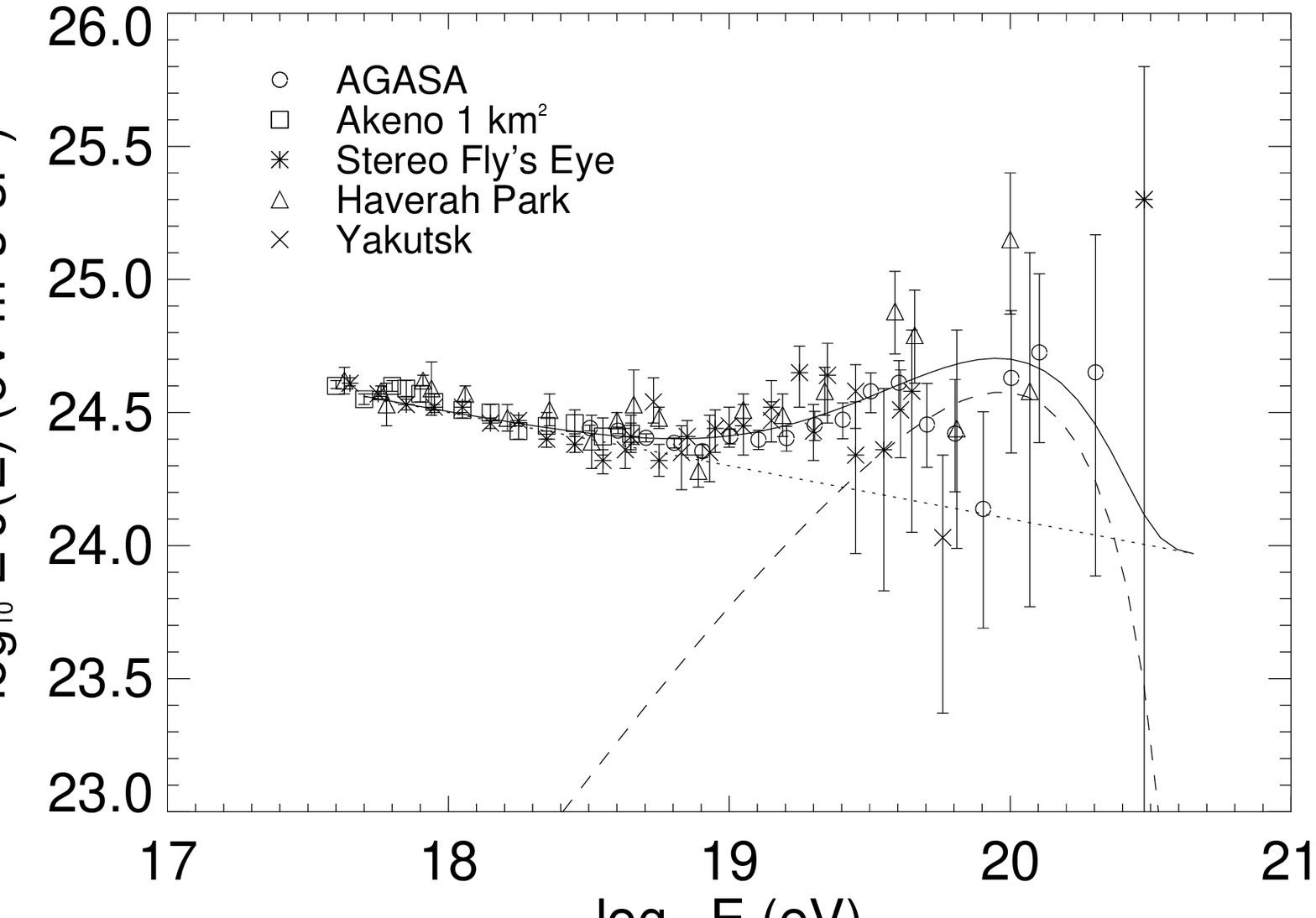}
  \bigskip\bigskip
  \caption{The best SM evolution fit to the cosmic ray data with a
  decaying particle mass of $10^{12}\gev$. The dotted line indicates
  the extrapolation of the power-law component from lower energies,
  while the dashed line shows the decay spectrum; the solid line is
  their sum.}
  \label{fig:SMfit} 
\end{figure}

\begin{figure}[htb]
  \epsfxsize\hsize\epsffile{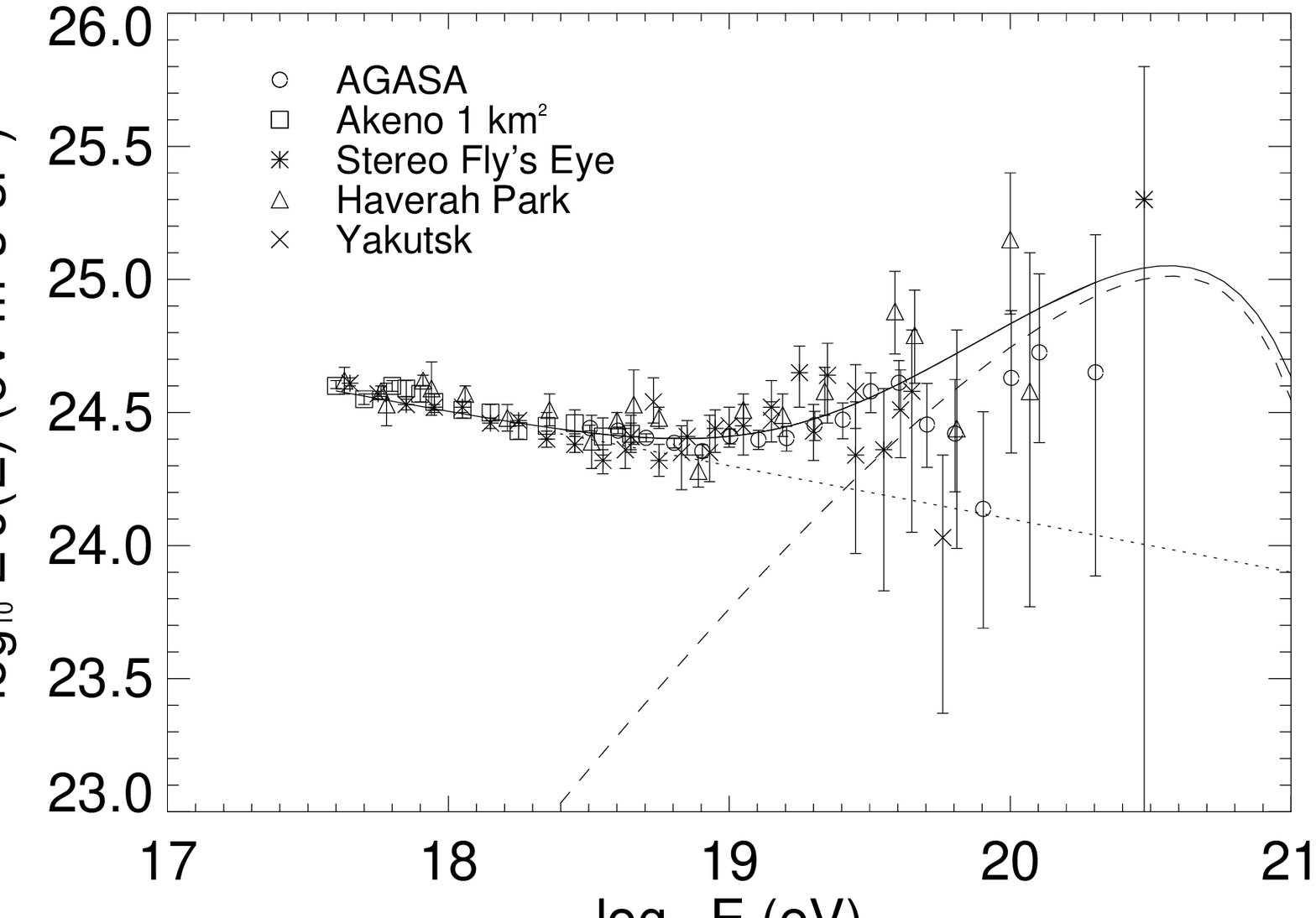}
  \bigskip\bigskip
  \caption{The best SUSY evolution fit to the cosmic ray data with a
  decaying particle mass of $5\times10^{12}\gev$. The dotted line
  indicates the extrapolation of the power-law component from lower
  energies, while the dashed line shows the decay spectrum; the solid
  line is their sum.}
  \label{fig:SUSYfit} 
\end{figure}

We have performed a least-squares fit to the data.
Figure~\ref{fig:SMfit} shows the best SM fit which corresponds to a
mass $M_X=10^{12}$~GeV, with $\chi^2=132$ for 81 degrees of
freedom. The low energy component (\ref{eq:TotalFlux}) and the total
spectrum are also shown. In calculating $\chi^2$ we ignore the highest
energy point at $\sim3.2\times10^{20}\ev$ which is based on a single
event, hence has a large flux uncertainty. Nevertheless this event is
important since it requires us to reject values of $M_X$ much below
$10^{12}\gev$ which cannot generate such an event, although they allow
a reasonable fit to the rest of data. In Fig.~\ref{fig:SUSYfit} we
plot the best SUSY fit taking a common mass of $M_\SUSY=400\gev$ for
all spartons; the favoured decaying particle mass is now
$M_X=5\times10^{12}\gev$, with a slightly lower $\chi^2$ of 130.

The assumption of 2-body decay may be rather naive for a superheavy
particle like a crypton which is expected to decay through very
high-order non-renormalisable operators \cite{crypton}. Many-body
decay distributes the total energy $M_X$ among several particles and
thus flattens the spectrum. We assume that many-body effects are
purely kinematical and hence can be encapsulated in the phase space of
the decay. In Appendix~\ref{MultiBody} we calculate $\rho_n(z)$, the
probability density that one parton carries off a fraction $z$ of the
total available energy per parton $M_X/2$. For $n\ge3$ we get
\begin{equation}
  \label{eq:rho3}
  \rho_n(z) \propto z (1-z)^{n-3}.
\end{equation}
To leading order in $\alpha_s$ the particle flux is then given by
\begin{equation}
 \label{eq:E3FluxMultiBody}
  E^3 J^\halo (E)= B x^3 \int^1_x \frac{\rmd z}{z} 
  \rho_n\left(\frac{x}{z}\right)  D^h (z, M^2_X).
\end{equation}
In particular if $D^h(x,M^2_X)\propto\,(1-x)^{a(M^2_X)}$ as
$x\rightarrow1$, the differential particle flux decreases as
$J^\halo(x)\propto(1-x)^{a(M^2_X)+n-2}$.

\begin{figure}[htb]
  \epsfxsize\hsize\epsffile{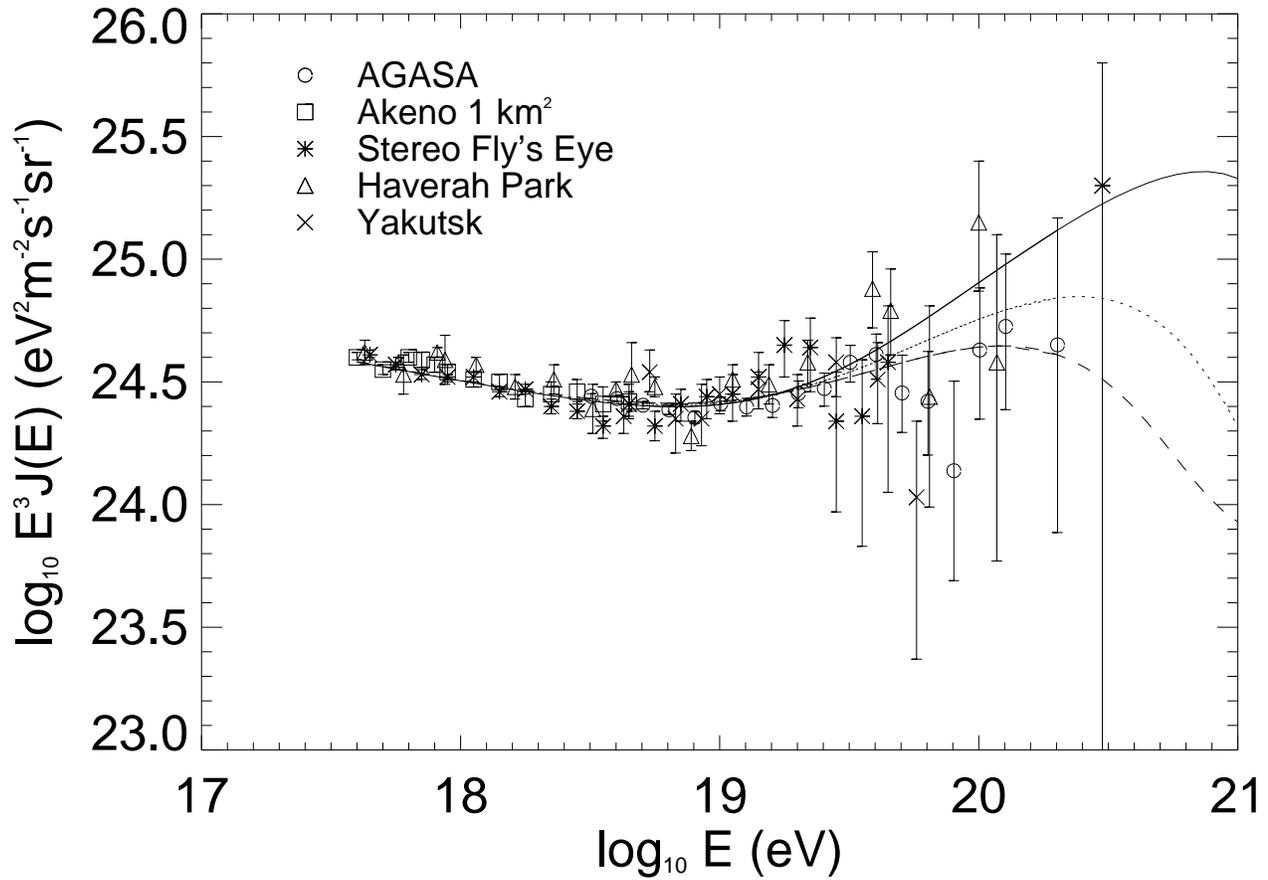}    
  \bigskip\bigskip
  \caption{Cosmic ray data compared with SUSY evolved spectra for a
  decaying particle mass of $10^{13}\gev$ with $M_\SUSY=400\,\gev$,
  and many-body decays to $n$ partons: $n=2$ (solid line), $n=8$
  (dotted line) and $n=16$ (dashed line).}  
  \label{fig:nbody}
\end{figure}

For many-body $X$ decays the total flux is
\begin{equation}
  \label{eq:TotalFluxMultiBody}
   E^3 J(E) = \frac{k}{E^m} +
   B x^3 \int^1_x \frac{\rmd z}{z} \rho_n\left(\frac{x}{z}\right) 
   D^\baryon (z , M^2_X),
\end{equation}
where $n$ is the number of partons into which $X$ decays. In
Fig.~\ref{fig:nbody} we plot SUSY spectra with $M_X=10^{13}\gev$,
$M_\SUSY=400\,\gev$ and $n=2,8,16$. As $n$ increases the spectrum
flattens since the average momentum is pushed to lower values of $x$
(see Appendix~\ref{MultiBody}). For $n=2,8,16$ we get
$\chi^2=133,127,126$ respectively. Clearly many-body decays fit the
data better which is encouraging as this would be natural in the
crypton model \cite{crypton}. With the forthcoming improvement in
event statistics a joint fit to the decaying particle mass and decay
multiplicity would be warranted.

\section{Discussion}

We now compare these results to our previous work in which the
fragmentation spectra were found by directly running HERWIG at the
decaying particle mass scale \cite{bs98}, as well as to other work
where numerical solutions to the DGLAP evolution were also reported
\cite{Rubin,fk01}, and, finally, to a recently proposed independent
method for calculating the fragmentation spectrum \cite{bk01}.

First it should be emphasised that most studies of `top-down' models
have employed the MLLA approximation to the fragmentation functions
\cite{bbv98,tdgamnu,bkv97,bk98} which, as noted earlier, is valid only
at $x\ll1$ where a very small (and uncertain) amount of energy is
released, and is thus {\em incapable} of being normalised as was done
in these papers. Other studies \cite{td} have used a convenient (and
properly normalised) parameterisation of the fragmentation function
based on PETRA data at $\sim20\gev$ \cite{Hill} which, however, does
not account for the large scaling violations in evolving up to the
very high energies of interest here. Moreover in both these approaches
it was {\em assumed} that the ratio of pions to baryons in the cascade
remains fixed at its low energy value of 20:1 at all energies. (In
reality one would expect the FF of say neutrinos to evolve differently
from that of baryons due to new processes at high energies e.g. decays
of heavy flavours.) All these studies focussed on a decaying particle
mass of $\sim10^{13-16}\gev$, close to the GUT scale.

These difficulties were emphasised in our previous work \cite{bs98}
where HERWIG was used to make the first realistic estimate of the
fragmentation function for super-massive particle decay, including the
expected large scaling violations. It was shown that the UHECR data
favoured a decaying particle mass of ${\cal O}(10^{12})\gev$ --- close
to the `hidden sector' scale of SUSY breaking rather than to the GUT
scale. However this work suffered itself from the fact that HERWIG
overproduces baryons by a factor of $\sim3$ at high $x$ (see
Fig.~\ref{fig:LEP_p}); moreover it did not account for the effects of
supersymmetry on the development of the cascade. Both these issues
were addressed in Ref.\cite{Rubin} which suggested that the way
forward was to evolve measured fragmentation functions at LEP up to
the decaying particle mass. For the SUSY case, sparton FFs were
generated at $10^4\gev$ using the event generator PYTHIA \cite{pythia}
and the evolution done for a supergravity model with parameters
$M_0=800\gev$, $M_{1/2}=200\gev$, $A_0=0$, $\tan\beta=10$,
sgn($\mu$)=+. Apart from highlighting the high-$x$ baryon
overproduction problem in the previous work \cite{bs98} for the SM
case, this work showed that the effects of supersymmetry could be
important in that the most likely decaying particle mass increased
from $\sim10^{12}\gev$ for the SM to $\sim10^{13}\gev$ for SUSY
\cite{Rubin}. Subsequently the DGLAP equations were also solved in
Ref.\cite{fk01} following a similar procedure but these authors
favoured the possibility that the decaying particles are uniformly
distributed in intergalactic space rather than being clustered in the
halo. Then the decay spectrum undergoes GZK processing, developing a
`bump' below the putative cutoff; by fitting this to the
(unnormalised) data from AGASA, Fly's Eye and Haverah Park they found
an even higher decaying particle mass of $\sim10^{15}\gev$
\cite{fk01}. (For the physically better motivated case of a clustered
halo population, their best-fit mass is $10^{12}\gev$ in agreement
with previous work \cite{bs98,Rubin}, but since the FFs for this mass
were not presented, we cannot compare their results with ours in
detail here.) Finally in Ref.\cite{bk01} a new Monte Carlo simulation
for jet fragmentation was presented and used to calculate the SUSY FFs
for a $10^{12}\gev$ mass particle; however the SM FFs were not
presented.\footnote{The authors acknowledge errors in their previous
analytic calculation of the SUSY FF in the MLLA approximation
\cite{bk98}, which was used in several studies of TD models
\cite{bbv98,tdgamnu}.}

\begin{figure}[htb]
  \epsfxsize\hsize\epsffile{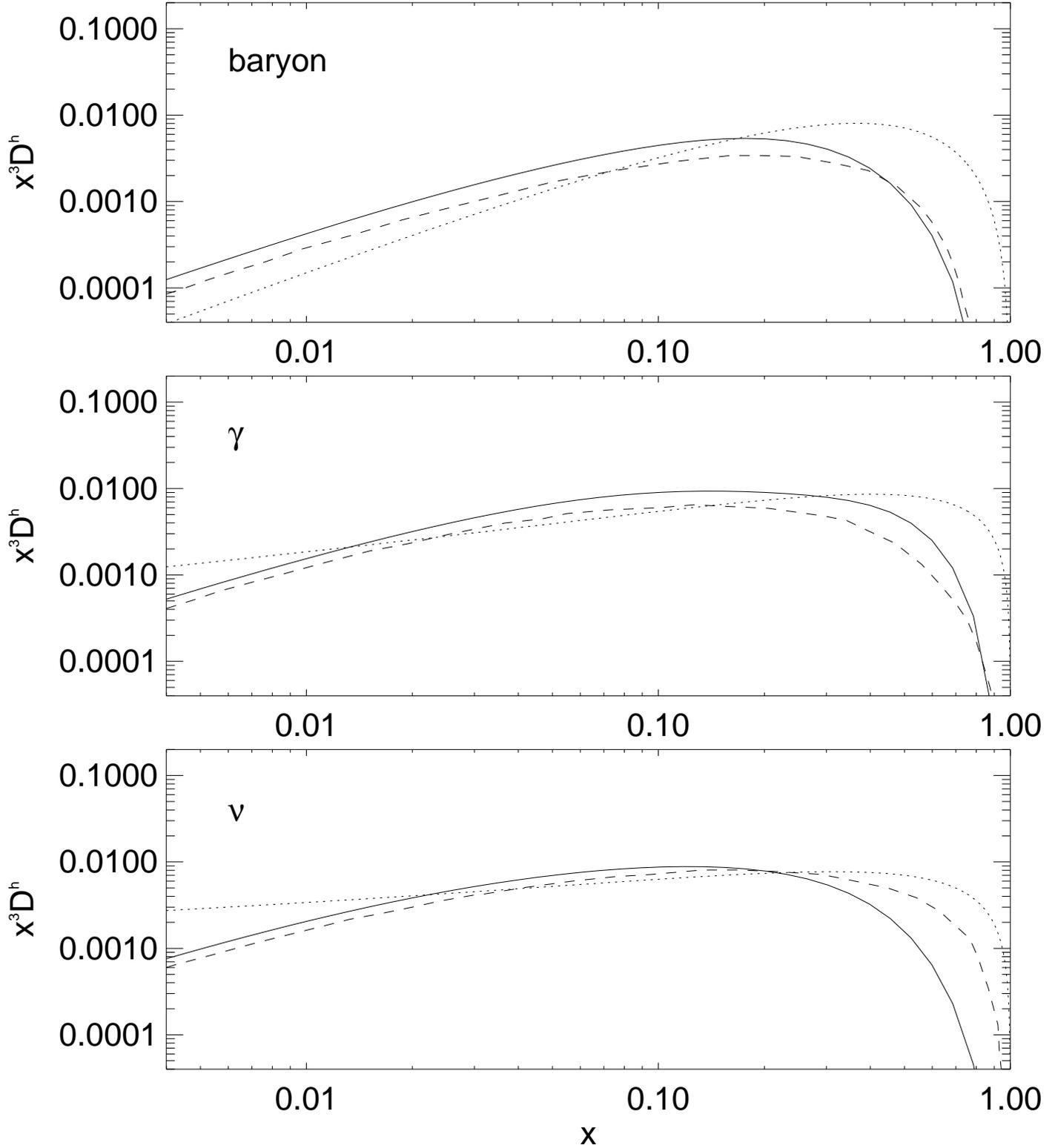} 
  \bigskip\bigskip
  \caption{Comparison of fragmentation functions of baryons (top
  panel), photons (middle panel) and neutrinos (bottom panel) obtained
  from (SM) DGLAP evolution in the present work (solid lines) and in
  Ref.\protect\cite{Rubin} (dashed lines), and using HERWIG
  \protect\cite{bs98} (dotted lines).}  
\label{fig:comp_SM}
\end{figure}

In Fig.~\ref{fig:comp_SM} we compare the (flavour-averaged) FFs of
baryons, photons and neutrinos for a $10^{12}\gev$ mass decaying
particle obtained from HERWIG \cite{bs98}, with the results from (SM)
DGLAP evolution in Ref.\cite{Rubin}, as well as in the present
study. The baryon overproduction problem at high $x$ is evident in the
HERWIG results; in fact photons and neutrinos now dominate at {\em
all} energies. Secondly the results obtained by DGLAP evolution agree
reasonably well, keeping in mind that Ref.\cite{Rubin} incorporated
gluon coherence \cite{ESW} and also next-to-leading order corrections
for both $\alpha_s$ evolution and the splitting functions, which we
have not done in the present work.

\begin{figure}[htb]
 \epsfxsize\hsize\epsffile{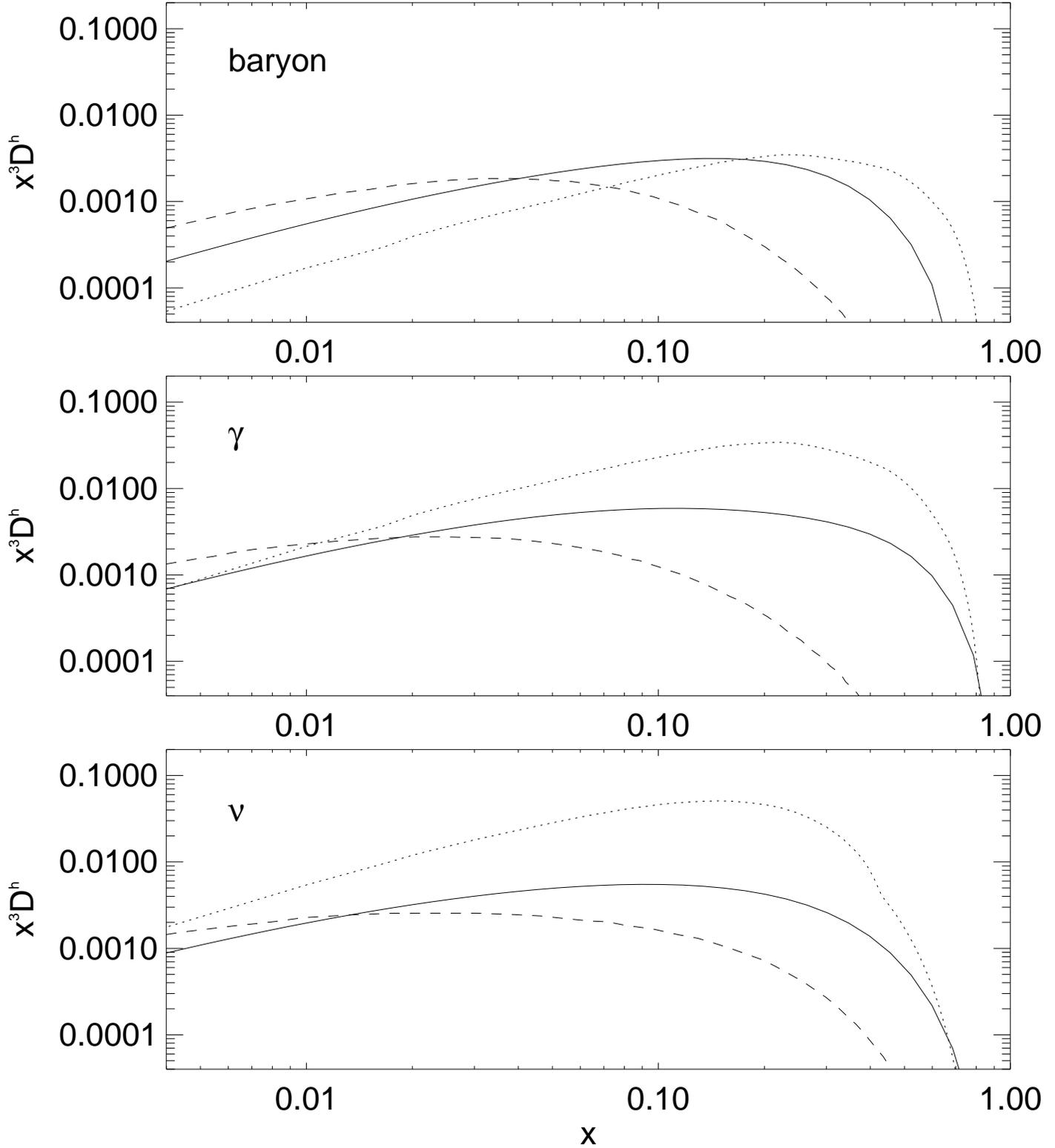} 
  \bigskip\bigskip
  \caption{Comparison of fragmentation functions of baryons (top
  panel), photons (middle panel) and neutrinos (bottom panel) obtained
  from (SUSY) DGLAP evolution in the present work (solid lines) and in
  Ref.\protect\cite{Rubin} (dashed lines), and using a new fragmentation code
  \protect\cite{bk01} (dotted lines).}  
  \label{fig:comp_SUSY}
\end{figure}

In Fig.~\ref{fig:comp_SUSY} we compare the (flavour-averaged) FFs for
baryons, photons and neutrinos obtained from (SUSY) DGLAP evolution in
Ref.\cite{Rubin} and in present work, with the results from the new
fragmentation code of Ref.\cite{bk01}. There are significant
differences between the results. Note that we have taken sparton FFs
to be {\em zero} at threshold rather than generating them with PYTHIA
above threshold as was done in Ref.\cite{Rubin}; this is presumably
the major reason for the difference in our results and reflects the
present degree of theoretical uncertainty in this approach. The
results obtained by the new Monte Carlo simulation for jet
fragmentation \cite{bk01} differ significantly from both DGLAP
evolution calculations. Since the SM FFs were not presented in this
work we cannot check whether the new code \cite{bk01} can reproduce
the (consistent) results for SM DGLAP evolution discussed above. In
view of these discrepancies, further work is clearly necessary to
calculate the spectrum reliably in the supersymmetric case.

In summary we have developed a new numerical code to solve the DGLAP
evolution equations for fragmentation functions in heavy particle
decay. We have used this to calculate the expected spectra of baryons,
photons and neutrinos from the decays of hypothetical metastable dark
matter particles of mass $10^{12}\gev$ clustered in the halo of our
Galaxy, both for the case of Standard Model evolution and for a simple
supersymmetric model. The shape of the fragmentation spectrum (of
either baryons or photons) fits rather well the new component of
ultra-high energy cosmic rays extending beyond the GZK energy; the fit
improves if the effects of supersymmetry and, more importantly, of
many-body decays are taken into account.

It is sometimes stated that the decaying particle model is already in
conflict with observations because the dominant particles in the
fragmentation cascades are predicted to be photons rather than baryons
which are favoured by experiment \cite{ahvwz00}. However as mentioned
earlier the photons may be significantly attenuated in their passage
to Earth due to interactions on the radio background in the halo,
particularly if there is a fossil radio `ghost' around our Galaxy
\cite{ghost}. Whether this is indeed possible, subject to the EGRET
bound on the low energy $\gamma$-rays which will be created by such
electromagnetic cascades, will be discussed by us elsewhere. Also of
interest is the expected flux of high energy neutrinos which can be
detected in forthcoming experiments such as ANTARES \cite{antares} and
ICECUBE \cite{icecube}. Most importantly the Pierre Auger Observatory
\cite{auger} will significantly increase the statistics of UHECRs
enabling a better measurement of the spectrum, as well as of the
composition and (an)isotropy. It will soon be possible to definitively
test the exciting possibility that cosmic rays have already shown us a
glimpse of physics far beyond the Standard Model.

\section{Acknowledgments} 

We are grateful to Peter Richardson and Mike Seymour for assistance in
the use of HERWIG and John Ellis, Neil Rubin, Alan Watson and Bryan
Webber for many discussions. We would also like to thank Zoltan Fodor,
Michael Kachelrie\ss, Sandor Katz and Motohiko Nagano for helpful
correspondence. R.T. is supported by a Marie Curie Fellowship
No.~HPMF-CT-1999-00268.

\appendix

\section{Many-body Decay Kinematics} \label{MultiBody}

We briefly review how to calculate the phase space $R_n$ for the decay
of $X$ into $n$ partons (see Ref.\cite{BycklingKajantie} for further
details); this is needed to calculate the probability density of
obtaining a decay parton with energy $zM_X/2$. 

\begin{figure}[htb]
  \epsfxsize\hsize\epsffile{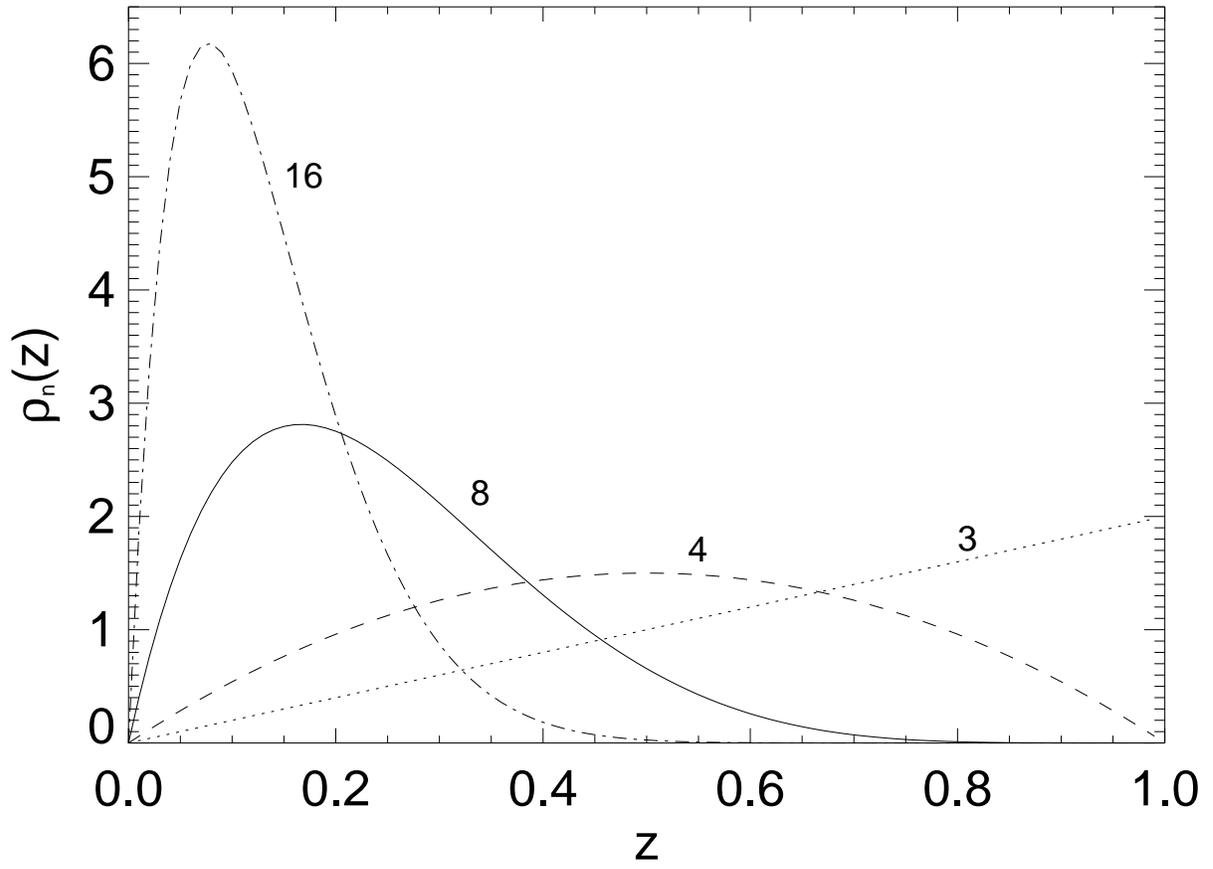}
  \bigskip\bigskip
\caption{Final state phase space probability density
  $\rho_n(z)$ for $n=3,4,8,16$.}
  \label{fig:rho}
\end{figure}

From a kinematical point of view the decay $X\rightarrow a_1+a_2+a_3+
\dots +a_n$ can be decomposed recursively as a 2-body decay
$X\rightarrow a_1+A_1$ where $A_1$ is a particle with 4-momentum equal
to the sum of 4-momenta of particles $a_2,\, a_3 \dots a_n$.  Then
$A_1$ is decayed into two particles $A_1\rightarrow a_2+A_2$ where
$A_2$ has as 4-momentum the sum of the 4-momenta of $a_3 \dots a_n$,
and so on up to the decay $A_{n-2}\rightarrow a_{n-1}+a_n$. This
recursive procedure allows the phase space integral $R_n(M_X,m_1,\dots
m_n)$, $n\ge 3$, to be written as
\begin{eqnarray}
 \nonumber
 R_n &=& \int^{(M_X-m_1)^2}_{(m_2+m_3+\dots+m_n)^2}\frac{\rmd M_1^2}{M_1^2}
 R_2(M_X,m_1,M_1)
 \int^{(M_1-m_2)^2}_{(m_3+\dots+m_n)^2} \frac{\rmd M_2^2}{M_2^2}
 R_2(M_1,m_2,M_2) \\  \label{eq:nDecay} && \dots
 \int^{(M_{n-3}-m_{n-2})^2}_{(m_{n-1}+m_n)^2} \frac{\rmd M_{n-2}^2}{M_{n-2}^2}
 R_2(M_{n-3},m_{n-2},M_{n-2}) \times R_2(M_{n-2},m_{n-1},m_n),
\end{eqnarray}
where $M_i$ is the invariant mass of $A_i$ and $R_2(m_a,m_b,m_c)$ is
the phase space integral for the 2-body decay $a\rightarrow b+c$:
\begin{eqnarray}
  \label{eq:TwoBody}
  R_2 &=&\frac{1}{8m_a} \lambda^{1/2}(m_a^2,m_b^2,m_c^2) \int\rmd\Omega, \\
  \lambda(x,y,z) &\equiv& x^2+y^2+z^2-2xy-2yz-2xz.
\end{eqnarray}
We shall study massless partons for which Eq.~(\ref{eq:nDecay}) reduces to
\begin{eqnarray}
  \nonumber
  R_n &\propto& \int^{M_X^2}_0 \frac{\rmd M_1^2}{M_1^2}(M_X^2-M_1^2)
  \int^{M_1^2}_0 \frac{\rmd M_2^2}{M_2^2}(M_1^2-M_2^2) \\ && \dots 
  \label{eq:nDecayMassless}
  \int^{M_{n-3}^2}_0 \frac{\rmd M_{n-2}^2}{M_{n-2}^2}(M_{n-3}^2-M_{n-2}^2)
  M_{n-2}^2.
\end{eqnarray}
Let $\rho_n(z)$ be the probability density to get parton $a_i$ with an
energy fraction $z=2E_i/M_X$ ($M_X/2$ is the maximum energy that a
parton can carry, whatever $n$ is). By symmetry this probability is
independent of the parton considered.  Let us choose parton
$a_1$. Then $z=1-M_1^2/M_X^2$ and
\begin{eqnarray}
  \nonumber
  \rho_n(z)dz &\propto& \frac{\rmd M_1^2}{M_1^2}(M_X^2-M_1^2)
  \int^{M_1^2}_0 \frac{\rmd M_2^2}{M_2^2}(M_1^2-M_2^2) \\ && \dots 
  \label{eq:rho}
  \int^{M_{n-3}^2}_0 \frac{\rmd M_{n-2}^2}{M_{n-2}^2}(M_{n-3}^2-M_{n-2}^2)
  M_{n-2}^2.
\end{eqnarray}
The multiple integration can easily be done to obtain
\begin{equation}
  \label{eq:rho2}
  \rho_n(z)\rmd z \propto (M_X^2-M_1^2)(M_1^2)^{n-3}\rmd M_1^2
  \propto z(1-z)^{n-3}\rmd z.
\end{equation}
After normalisation we finally obtain:
\begin{eqnarray}
  \label{eq:rhoFinal}
  \rho_2(z) &=& \delta(1-z), \\
  \rho_n(z) &=& (n-1)(n-2)z(1-z)^{n-3}, \quad n\ge 3.
\end{eqnarray}
We plot this distribution in Fig.~\ref{fig:rho}. As the number of
final particles $n$ increases, the total energy $M_X$ has to be shared
between more particles and consequently less energy goes to each one
of them. The probability density peaks at smaller values of $z$ as $n$
grows, accordingly the mean value of $z$ decreases as
$\left<z\right>_n=2/n$.


\end{document}